\documentclass[aps,pra,twocolumn,showpacs,preprintnumbers,amsmath,amssymb,superscriptaddress]{revtex4-2}

\usepackage{graphicx}
\usepackage{epsfig}
\usepackage{amsmath}
\usepackage{amssymb}
\usepackage{dcolumn}
\usepackage{bm}
\usepackage{braket}
\usepackage{mathtools}
\usepackage[nice]{nicefrac}
\usepackage{bbm}
\usepackage[usenames,dvipsnames]{xcolor}
\usepackage[colorlinks=true,citecolor=MidnightBlue,linkcolor=MidnightBlue,urlcolor=MidnightBlue]{hyperref}
\usepackage{siunitx}

\definecolor{Light Salmon Pink}{RGB}{255, 154, 162}
\definecolor{Melon}{RGB}{255, 183, 178}
\definecolor{Very Pale Orange}{RGB}{255, 218, 193}
\definecolor{Dirty White}{RGB}{226, 240, 203}
\definecolor{Magic Mint}{RGB}{181, 234, 215}
\definecolor{Crayola's Periwinkle}{RGB}{199, 206, 234}
\definecolor{Dark Pastel Red}{RGB}{194, 59, 35}
\definecolor{Carrot Orange}{RGB}{243, 154, 39}
\definecolor{Minion Yellow}{RGB}{234, 218, 82}
\definecolor{Dark Pastel Green}{RGB}{3, 192, 60}
\definecolor{Silver Lake Blue}{RGB}{87, 154, 190}
\definecolor{Dark Pastel Purple}{RGB}{151, 110, 215}

\definecolor{matblue}{RGB}{0, 113.9850, 188.9550}
\definecolor{matred}{RGB}{216.7500, 82.8750, 24.9900}
\definecolor{matyellow}{RGB}{236.8950, 176.9700, 31.8750}
\definecolor{matpurple}{RGB}{125.9700, 46.9200, 141.7800}
\definecolor{matgreen}{RGB}{118.8300, 171.8700, 47.9400}
\definecolor{matcyan}{RGB}{76.7550, 189.9750, 237.9150}

\PassOptionsToPackage{numbers,sort&compress}{natbib}


\renewcommand{\BibitemShut}[1]{} 

\colorlet{BLUE}{blue}

\newcommand{\op}[1]{\hat{#1}} 

\begin{document}

\newcommand*{\MAINZ}{QUANTUM, Institut f\"ur Physik, Johannes Gutenberg-Universit\"at Mainz, Staudingerweg 7, 55128 Mainz, Germany}
\newcommand*{\ERLANGEN}{Institut f\"ur Optik, Information und Photonik, Friedrich-Alexander Universit\"at Erlangen-N\"urnberg, Staudtstr. 1, 91058 Erlangen, Germany}
\newcommand*{\SAOT}{Erlangen Graduate School in Advanced Optical Technologies (SAOT), Friedrich-Alexander Universit\"at Erlangen-N\"urnberg, Paul-Gordan-Str. 6, 91052 Erlangen, Germany}
\homepage{http://www.quantenbit.de}

\title{Phase Retrieval in Incoherent Diffractive Imaging using higher-order photon correlation functions}

\author{M.~Bojer}
\affiliation{Friedrich-Alexander-Universität Erlangen-Nürnberg, Quantum Optics and Quantum Information, Staudtstr. 1, 91058 Erlangen, Germany}
\author{J.~Eckert}
\affiliation{Friedrich-Alexander-Universität Erlangen-Nürnberg, Quantum Optics and Quantum Information, Staudtstr. 1, 91058 Erlangen, Germany}
\author{S.~Karl}
\affiliation{Friedrich-Alexander-Universität Erlangen-Nürnberg, Quantum Optics and Quantum Information, Staudtstr. 1, 91058 Erlangen, Germany}
\author{S.~Richter}
\affiliation{Friedrich-Alexander-Universität Erlangen-Nürnberg, Quantum Optics and Quantum Information, Staudtstr. 1, 91058 Erlangen, Germany}
\author{J.~von~Zanthier}
\affiliation{Friedrich-Alexander-Universität Erlangen-Nürnberg, Quantum Optics and Quantum Information, Staudtstr. 1, 91058 Erlangen, Germany}

\date{\today}

\begin{abstract}
\noindent To obtain spatial information about an arbitrary object in x-ray structure analysis, the standard method is to measure the intensity in the far field, i.e., the first-order photon correlation function of the coherently scattered x-ray photons (coherent diffractive imaging). Recently, it was suggested to record alternatively the incoherently scattered photons and measure the second-order photon correlation function to reconstruct the geometry of the unknown object (incoherent diffractive imaging). Yet, besides various advantages of the latter method, both techniques suffer from the so-called phase retrieval problem. Lately, an ab-initio phase retrieval algorithm to reconstruct the phase of the so-called structure factor of the scattering objects based on the third-order photon correlation function was reported. The algorithm makes use of the so-called closure phase, which contains important, yet incomplete phase information, well-known from triple correlations and their bispectrum in speckle masking and astronomy applications. Here, we provide a detailed analysis of the underlying scheme and quantities in the context of x-ray structure analysis. In particular, we explicitly calculate the third-order photon correlation function in a full quantum mechanical treatment and discuss the uniqueness of the closure phase equations constructed from it. In this context, we recapitulate the sign problem of the closure phase and how it can be lifted using redundant information. We further show how the algorithm can be improved using even higher-order photon correlation functions, e.g., the fourth-order correlation function, delivering new phase relations appearing in the four-point correlations. 
\end{abstract}

\maketitle

\section{Introduction}
\noindent The goal of so-called coherent diffractive imaging (CDI) in x-ray structure analysis is to reconstruct the geometry of an object via the detection of the intensity of the coherently scattered x-ray photons in the far field. The detected intensity distribution is thereby proportional to the modulus squared of the so-called structure factor, the Fourier transform of the real space distribution of the object. To reconstruct the object via an inverse Fourier transform, the phase of the structure factor is thus needed. However, since only the intensity of the scattered radiation is measured, the phase information is missing. This is the so-called phase problem of CDI. Several so-called phase retrieval algorithms have been developed over the last years to recover the phase when only the intensity of the scattered light is known~\cite{Fienup78,Fienup82,Kowalczyk90,Belenkii04,Shechtman2014,Strekalov14,Shechtman2015,Cao16,Wang18,nakasako2020methods}.\\
Recently, it was suggested to measure the incoherently scattered light, which for CDI is usually considered to be detrimental since it leads merely to a constant background, in order to reconstruct the geometry of the object via so-called incoherent diffractive imaging (IDI)~\cite{Classen2017,schneider2018quantum,Trost_2020}. The idea is to use the second-order photon correlation function instead of the intensity, i.e., the first-order photon correlation function. It can be shown that the second-order correlation function of the incoherently scattered photons equally gives access to the modulus squared of the structure factor. Thus, IDI, equally to CDI, requires the phase of the structure factor to be determined in an independent manner. It is well-known from speckle masking and astronomy applications that triple correlations and their bispectrum contain some useful yet incomplete phase information about the structure factor via the so-called closure phase~\cite{gamo1963triple,Lohmann1983,Lohmann1984,Cornwell1989,Kang1991,Marathay1994,Ofir2006T,Ofir2006A,Malvimat2013,Dravins2013,Nunez2015,Shoulga2017}. However, since only the cosine of the closure phase is available, a sign ambiguity arises. In that context, an ab-initio phase retrieval algorithm based on the third-order photon correlation function was recently proposed~\cite{Peard23}, which demonstrates how to get rid of the sign ambiguity using redundant information. Here, we analyse the sign ambiguity problem in more detail and show how even higher-order photon correlation functions, e.g., the fourth-order photon correlation function, can be used to further reduce the ambiguities.\\
The paper is organized as follows: In Sec.~\ref{sec:Structure_Factor}, we introduce the so-called structure factor, the Fourier transform of the real space distribution of the object. In Sec.~\ref{sec:Coherent_Imaging}, we recapitulate the basic working principle of coherent diffractive imaging and explain the phase problem. In Sec.~\ref{sec:Incoherent_Imaging}, we first introduce the so-called complex degree of coherence and show its relation to the structure factor. We then explain in detail how the second- and third-order photon correlation functions can be used to reconstruct the missing phase via the so-called closure phase. In this context, we analyse the sign ambiguity of the closure phase and recapitulate the proposed algorithm of Ref.~\cite{Peard23}. We end this section by demonstrating how the new phase relations occurring for four-point correlations in the fourth-order photon correlation function can be used to further reduce the previously discussed ambiguity. We summarise our results and finally conclude in Sec.~\ref{sec:Conclusion}.

\section{Structure factor}
\label{sec:Structure_Factor}
\noindent In order to model the structure of a complex object, we consider $N$ point-like atoms, assumed to be two-level systems, located at positions $\bm{R}_i$, $i\in\{1,...,N\}$. The real space distribution is thus given by
\begin{equation}
	\label{eq:real_space_dist}
	S(\bm{r}) = \sum_{i=1}^N \delta(\bm{r}-\bm{R}_i)\,.
\end{equation}
The Fourier transform of the real space distribution is the so-called structure factor denoted by
\begin{align}
	\tilde{S}(\bm{q}) &= \int d^3r\,S(\bm{r})e^{-i\bm{q}\bm{r}}\nonumber\\
	&= \sum_{i=1}^N \int d^3r\,\delta(\bm{r}-\bm{R}_i)e^{-i\bm{q}\bm{r}}= \sum_{i=1}^N e^{-i\bm{q}\bm{R}_i}\,.
\end{align}
We can write the structure factor in terms of its magnitude and phase as
\begin{equation}
	\tilde{S}(\bm{q})=|\tilde{S}(\bm{q})|e^{i\phi(\bm{q})}\,,
\end{equation}
which makes it evident that both the magnitude and phase are needed to reconstruct the real space distribution via an inverse Fourier transform
\begin{equation}
    S(\bm{r}) = \int d^3q\,|\tilde{S}(\bm{q})|e^{i\phi(\bm{q})} e^{i\bm{q}\bm{r}}\,.
\end{equation}
Now, assuming a real-valued real space distribution $S(\bm{r})\in \mathbb{R}$, we find
\begin{align}
	\tilde{S}(-\bm{q})&=|\tilde{S}(-\bm{q})|e^{i\phi(-\bm{q})}=\int d^3r\,S(\bm{r})e^{i\bm{q}\bm{r}}\nonumber\\
	&=\tilde{S}^*(\bm{q})=|\tilde{S}(\bm{q})|e^{-i\phi(\bm{q})}
\end{align}
and thus
\begin{equation}
	|\tilde{S}(-\bm{q})|=|\tilde{S}(\bm{q})|\,,\quad \phi(-\bm{q})=-\phi(\bm{q})\,,
\end{equation}
from which, in particular, $\phi(0)=-\phi(0)=0$ follows. By writing Eq.~\eqref{eq:real_space_dist}, we implicitly defined an origin located at some point $\bm{R}$. Let us assume that we would have chosen a different origin at $\bm{R}'$ with $\bm{R}'=\bm{R}+\Delta\bm{R}$. With respect to the new origin, the real space distribution reads
\begin{equation}
	S'(\bm{r}) = \sum_{i=1}^N \delta(\bm{r}-\bm{R}'_i) = \sum_{i=1}^N \delta(\bm{r}-\bm{R}_i + \Delta\bm{R})=S(\bm{r}+\Delta\bm{R})\,,
\end{equation}
where we used $\bm{R}'_i =\bm{R}_i - \Delta\bm{R}$. That is, we obtain a shifted real space distribution. Calculating the structure factor of the shifted object leads to
\begin{align}
	\label{eq:shift_dist}
	\tilde{S}'(\bm{q})&=|\tilde{S}'(\bm{q})|e^{i\phi'(\bm{q})}=\int d^3r\,S(\bm{r}+\Delta\bm{R})e^{-i\bm{q}\bm{r}}\nonumber\\
	&= e^{i\bm{q}\Delta\bm{R}}\tilde{S}(\bm{q}) = |\tilde{S}(\bm{q})|e^{i[\phi(\bm{q})+\bm{q}\Delta\bm{R}]}\,.
\end{align}
Thus, the new magnitude and phase are given by $|\tilde{S}'(\bm{q})|=|\tilde{S}(\bm{q})|$ and $\phi'(\bm{q})=\phi(\bm{q})+\bm{q}\Delta\bm{R}$, i.e., the shift of the origin leads to a linear shift in the phase. We will come to this outcome later when we discuss the phase retrieval algorithm based on higher-order photon correlation functions. We end this section by noting that the aim of x-ray structure analysis is to determine the real space distribution $S(\bm{r})$ via scattered x-ray photons.

\section{Coherent diffractive imaging}
\label{sec:Coherent_Imaging}
\noindent In this section, we recall some basic concepts of CDI, where it is assumed that the incoming photons are coherently scattered by the atoms. The outgoing electric field can be written as a superposition of the incoming field and a scattered spherical wave field, which is proportional to the so-called scattering factor $f(\bm{q})$, where $\bm{q}$ denotes the momentum transfer of the scattering process. The scattering factor is thus proportional to the structure factor
\begin{equation}
	f(\bm{q}) \propto \tilde{S}(\bm{q})=\int d^3r\,S(\bm{r})e^{-i\bm{q}\bm{r}}\,.
\end{equation} 
Since the energy of the photons is unchanged in a coherent scattering process, the momentum transfer vectors $\bm{q}$ are confined to the so-called Ewald sphere. Therefore, to get full 3D information via detection of photons on a pixelated detector, the object has to be rotated. Moreover, to determine the real space distribution $S(\bm{r})$ via a Fourier transform of $\tilde{S}(\bm{q})$, full information about the complex function $\tilde{S}(\bm{q})$ is needed, i.e., the modulus and the phase have to be known. However, the measured intensity of the coherently scattered photons is merely proportional to $|f(\bm{q})|^2\propto |\tilde{S}(\bm{q})|^2$. That is, one has access to the absolute value, but not to the phase of the scattering factor. This is the so-called phase problem of CDI. Many phase retrieval algorithms have been developed to reconstruct the scattering phases, e.g., where reasonable constraints are imprinted by use of suitable projections~\cite{Fienup78,Fienup82,Kowalczyk90,Strekalov14}, via Cauchy-Riemann approaches using the Cauchy-Riemann equations~\cite{Belenkii04}, ptychographical methods~\cite{Cao16,Wang18}, or so-called greedy methods~\cite{Shechtman2014}. In the following, we consider and extend a phase retrieval algorithm based on the third-order photon correlation function using incoherently scattered photons~\cite{Peard23}.

\section{Incoherent diffractive imaging}
\label{sec:Incoherent_Imaging}
\noindent In this section, we consider the case of IDI, where the incoherently emitted fluorescence light of the atoms is measured in the far field and the second-order photon correlation function is computed. We consider again $N$ two-level atoms with excited state $\ket{e}$ and ground state $\ket{g}$. Neglecting the dipole radiation pattern and a general constant prefactor, which would provide a meaningful physical unit, the positive electric field operator at time $t=0$ is given by~\cite{Agarwal1974}
\begin{equation}
	\op{E}^{(+)}(\bm{r}) = \sum_{i=1}^{N} e^{-i k_0 \hat{\bm{r}} \bm{R}_i} \op{S}_-^{(i)}\,.
\end{equation}
Here, $k_0=2\pi/\lambda$ denotes the wave number, $\hat{\bm{r}}$ the unit vector in direction of the observation, $\bm{R}_i$ the position of the $i$th atom, and $\op{S}_-^{(i)}$ the pseudo-spin lowering operator of the $i$th atom. For convenience, we define $\bm{k} \coloneqq k_0 \hat{\bm{r}}$. In this way, we can write the positive electric field operator as a function of $\bm{k}$ as
\begin{equation}
	\label{eq:Epos}
	\op{E}^{(+)}(\bm{k}) = \sum_{i=1}^{N} e^{-i \bm{k} \bm{R}_i} \op{S}_-^{(i)}\,.
\end{equation}
We start by calculating the far field intensity produced by the atoms, which can be written as~\cite{Glauber1963}
\begin{align}
	I(\bm{k})&= G^{(1)}(\bm{k},\bm{k}) = \braket{\op{E}^{(-)}(\bm{k})\op{E}^{(+)}(\bm{k})}\nonumber\\
	&= \sum_{i,j=1}^N e^{i\bm{k}\bm{R}_i} e^{-i\bm{k}\bm{R}_j}\braket{\op{S}_+^{(i)} \op{S}_-^{(j)}}\,.
\end{align}
Assuming the atoms to be initially in the fully excited state end thus emit the photons incoherently, the expectation value $\braket{\op{S}_+^{(i)} \op{S}_-^{(j)}}$ is zero unless $i=j$, whereas $\braket{\op{S}_+^{(i)} \op{S}_-^{(i)}}=1\,\forall i$. The intensity is then given by
\begin{equation}
	I(\bm{k}) = \sum_{i=1}^N \braket{\op{S}_+^{(i)} \op{S}_-^{(i)}} = N\,,
\end{equation}
which is a constant and thus gives no phase information about the real space distribution. However, in contrast to the intensity, the complex degree of coherence (CDC) $g^{(1)}(\bm{k}_1,\bm{k}_2)$, i.e., the cross-correlation between the electric fields in the directions $\bm{k}_1$ and $\bm{k}_2$ does reveal structure information since
\begin{align}
	\label{eq:CDC}
	g^{(1)}(\bm{k}_1,\bm{k}_2) &= \frac{G^{(1)}(\bm{k}_1,\bm{k}_2)}{\sqrt{G^{(1)}(\bm{k}_1,\bm{k}_1)}\sqrt{G^{(1)}(\bm{k}_2,\bm{k}_2)}}\nonumber\\
	&=\frac{1}{N}\sum_{i=1}^N e^{-i\bm{q}\bm{R}_i} = \frac{1}{N} \tilde{S}(\bm{q})=\frac{\tilde{S}(\bm{q})}{\tilde{S}(0)}\,,
\end{align}
where $\bm{q}=\bm{k}_2-\bm{k}_1$ is the momentum difference vector. As found in Eq.~\eqref{eq:CDC}, measuring the CDC of the emitted fluorescence light of the $N$ atoms would give full information about the real space distribution $S(\bm{r})$. However, determining experimentally the CDC is usually very hard~\cite{Classen2017}. Instead, we investigate the possibility of using higher-order photon correlation functions to get access to the structure factor when measuring incoherent fluorescence light in the following. 
As we show in App.~\ref{sec:AppA}, the second-order photon correlation function calculates to
\begin{align}
	g^{(2)}(\bm{k}_1,\bm{k}_2)=1-\frac{2}{N}+\frac{1}{N^2}|\tilde{S}(\bm{q})|^2\,.
 \label{eq:g2}
\end{align}
According to Eq.~\eqref{eq:g2}, $g^{(2)}(\bm{k}_1,\bm{k}_2)$ also gives access to the modulus of the structure factor, however not to the phase.
\section{Phase retrieval using higher-order photon correlation functions}
\noindent In order to get access to the phase of the structure factor $\tilde{S}(\bm{q})$, one can consider the phase relations appearing in the third-order photon correlation function. We start indeed by considering the third-order correlation function, however, will subsequently also investigate the fourth-order photon correlation function for this purpose. 
\subsection{Third-order photon correlation function}
\noindent The third-order photon correlation function calculates to (see App.~\ref{sec:AppB})
\begin{widetext}
	\begin{align}
		g^{(3)}(\bm{k}_1,\bm{k}_2,\bm{k}_3)&=1-\frac{6}{N}+\frac{12}{N^2}+\frac{N-4}{N^3}[|\tilde{S}(\bm{q}_1)|^2+|\tilde{S}(\bm{q}_2)|^2+|\tilde{S}(\bm{q}_1+\bm{q}_2)|^2]\nonumber\\
		&+ \frac{2}{N^3}|\tilde{S}(\bm{q}_1)||\tilde{S}(\bm{q}_2)||\tilde{S}(\bm{q}_1+\bm{q}_2)| \cos[\phi(\bm{q}_1+\bm{q}_2)-\phi(\bm{q}_1)-\phi(\bm{q}_2)]\,,
		\label{eq:g3main}
	\end{align}
\end{widetext}
where we denote $\bm{q}_1 = \bm{k}_2-\bm{k}_1$ and $\bm{q}_2 = \bm{k}_3-\bm{k}_2$. As can be seen from Eq.~\eqref{eq:g3main}, the third-order correlation function involves the so-called closure phase
\begin{equation}
	\Phi(\bm{q}_1,\bm{q}_2) \coloneqq \phi(\bm{q}_1+\bm{q}_2) - \phi(\bm{q}_1) - \phi(\bm{q}_2)\,,
\end{equation}
which is well-known also in astronomy. Since both the number of atoms $N$ as well as the magnitude of the structure factor $|\tilde{S}(\bm{q})|$ can be determined by the second-order photon correlation function, we can focus on the closure phase in the following. We see that Eq.~\eqref{eq:g3main} gives actually only access to the cosine of the closure phase, leading to a sign (and a $2\pi$ shift) ambiguity. However, as recently shown, the sign ambiguity can be lifted in certain cases, such that a complete phase retrieval using the third-order photon correlation function is possible. We will present a detailed analysis of the number of non-trivial inequivalent equations and the number of unknowns in the following and review the possibilities to unlift the sign ambiguity. We then show how the algorithm can be improved by additionally using the fourth-order photon correlation function.
\subsection{Uniqueness of closure phase equations and sign ambiguity lifting}
\noindent In the linear regime, i.e., in the far-field, the vectors $\bm{q}_1$ and $\bm{q}_2$ can be considered as difference vectors between different positions of pixels on a camera. Therefore, we will adapt our notation in the following and write
\begin{equation}
	\Phi(\bm{m},\bm{n}) = \phi(\bm{m}+\bm{n}) - \phi(\bm{m}) - \phi(\bm{n})
\end{equation}
with discrete difference vectors $\bm{m}$ and $\bm{n}$ with natural number entries. We now check the uniqueness of these equations for two different scenarios, namely a pixelated camera in one dimension and two dimensions, respectively.\\
\noindent Let us start by assuming that we have a camera of $M$ pixels along one dimension. Without loss of generality, we can set the first pixel of the triple correlations to be the first pixel of the camera. Now, we consider an arbitrary pixel defined by the difference value $u \in \{0,...,M-1\}$ with respect to the first pixel. The question now is, how many possibilities do we have to achieve $m+n=u$ with $m,n\leq u$ and $m,n\in \mathbb{N}$. There are exactly $u+1$ possibilities. This leads to a total number of equations given by
\begin{equation}
	\sum_{u=0}^{M-1} (u+1) = \sum_{u=1}^{M} u = \frac{1}{2} M (M+1)\,.
\end{equation}
However, not all of these equations are independent. In fact, all equations with $m$ and $n$ interchanged are equivalent since $\Phi(m,n) = \Phi(n,m)$ (redundant equations). In addition, $\phi(0)=0$, such that $\Phi(m,0)=\Phi(0,m)=0$ does not give additional information either (trivial equations). Let us first count the number of redundant equations, which for $M$ even reads $\frac{M^2}{4}$, whereas for $M$ odd is given by $\frac{M^2-1}{4}$. Further, the number of trivial equations is given by $M$. Therefore, the number of non-trivial inequivalent equations is given by
\small
\begin{align}
	F^{(1)}_\mathrm{even}(M) &\coloneqq \frac{1}{2} M (M+1) - \frac{1}{4} M^2 - M = \frac{1}{4} M (M-2)\,,\\
	F^{(1)}_\mathrm{odd}(M) &\coloneqq \frac{1}{2} M (M+1) - \frac{1}{4} (M^2-1) - M = \frac{1}{4} (M-1)^2\,.
\end{align}
\normalsize
In contrast, the number of unknown variables is $F^{(1)}_{\mathrm{even/odd}}(M) + M-2$, where $F^{(1)}_{\mathrm{even/odd}}(M)$ gives the number of closure phases $\Phi(m,n)$, i.e., essentially signs since the value of $|\Phi(m,n)|$ is known, and $M-2$ phases $\phi(m)$, namely $\phi(2),...,\phi(M-1)$ since $\phi(0)=0$ and $\phi(1)$ can be arbitrarily chosen, as it corresponds to a global shift, i.e., it defines the origin of the coordinate system [see Eq.~\eqref{eq:shift_dist}]. This means, in principle the number of unknowns is greater than the number of equations. However, since many of the unknowns are only signs, in certain cases it is possible to find a unique solution. To exemplify this statement, let us consider the example of a one-dimensional camera with 9 pixel. In this case, we have $F^{(1)}_\mathrm{odd}(9) = 16$ non-trivial inequivalent equations, namely
\begin{align}
	\label{eq:11}
	\Phi(1,1) &= \phi(2) - 2 \phi(1)\\
	\label{eq:21}
	\Phi(2,1) &= \phi(3) - \phi(2) - \phi(1)\\
	\label{eq:31}
	\Phi(3,1) &= \phi(4) - \phi(3) - \phi(1)\\
	\label{eq:41}
	\Phi(4,1) &= \phi(5) - \phi(4) - \phi(1)\\
	\label{eq:51}
	\Phi(5,1) &= \phi(6) - \phi(5) - \phi(1)\\
	\label{eq:61}
	\Phi(6,1) &= \phi(7) - \phi(6) - \phi(1)\\
	\label{eq:71}
	\Phi(7,1) &= \phi(8) - \phi(7) - \phi(1)\\
	\label{eq:22}
	\Phi(2,2) &= \phi(4) - 2 \phi(2)\\
	\label{eq:32}
	\Phi(3,2) &= \phi(5) - \phi(3) - \phi(2)\\
	\label{eq:42}
	\Phi(4,2) &= \phi(6) - \phi(4) - \phi(2)\\
	\label{eq:52}
	\Phi(5,2) &= \phi(7) - \phi(5) - \phi(2)\\
	\label{eq:62}
	\Phi(6,2) &= \phi(8) - \phi(6) - \phi(2)\\
	\label{eq:33}
	\Phi(3,3) &= \phi(6) - 2 \phi(3)\\
	\label{eq:43}
	\Phi(4,3) &= \phi(7) - \phi(4) - \phi(3)\\
	\label{eq:53}
	\Phi(5,3) &= \phi(8) - \phi(5) - \phi(3)\\
	\label{eq:44}
	\Phi(4,4) &= \phi(8) - 2 \phi(4) - \phi(1)\,,
\end{align}
but also $16$ unknown signs of the closure phases $\Phi(m,n)$ and $7$ phases $\phi(m)$, i.e., $\phi(2),...,\phi(8)$. We will show how to solve this problem later.\\
\noindent Let us now assume we have a two-dimensional camera of $M^2$ pixels. Without loss of generality, we can set the first pixel of the triple correlations to be the pixel at the left lower corner of the camera. Now, we consider a pixel defined by the difference vector $\bm{u}$ with respect to the first pixel. The question is, how many possibilities do we have to achieve $\bm{m}+\bm{n}=\bm{u}$ with $m_{x/y},n_{x/y}\leq u_{x/y}$ and $\bm{m},\bm{n}\in \mathbb{N}^2$. There are exactly $(u_x+1)(u_y+1)$ possibilities. This leads to a total number of equations given by
\begin{equation}
	\sum_{u_x,u_y=0}^{M-1} (u_x+1)(u_y+1) = \frac{1}{4} M^2 (M+1)^2\,.
\end{equation}
However, not all of these equations are independent. In fact, all equations with $\bm{m}$ and $\bm{n}$ interchanged are equivalent since $\Phi(\bm{m},\bm{n}) = \Phi(\bm{n},\bm{m})$ (redundant equations). In addition, $\phi(0)=0$, such that $\Phi(\bm{m},0)=\Phi(0,\bm{m})=0$ does not give additional information either (trivial equations). The number of redundant equations for $M$ even reads $\frac{1}{8} M^3 (M + 2)$, whereas for $M$ odd is given by $\frac{1}{8} (M-1) (M+1)^3$. Further, the number of trivial equations is given by $M^2$. Therefore, the number of non-trivial inequivalent equations is given by
\begin{widetext}
\begin{align}
	F^{(2)}_\mathrm{even}(M) &\coloneqq \frac{1}{4} M^2 (M+1)^2 - \frac{1}{8} M^3 (M + 2) - M^2 = \frac{1}{8} M^2 [M(M+2)-6]\,,\\
	F^{(2)}_\mathrm{odd}(M) &\coloneqq \frac{1}{4} M^2 (M+1)^2 - \frac{1}{8} (M-1) (M+1)^3 - M^2 = \frac{1}{8} (M-1)^2 [M(M+4)+1]\,.
\end{align}
\end{widetext}
In contrast, the number of unknown variables is $F^{(2)}_{\mathrm{even/odd}}(M) + M^2-3$, where $F^{(2)}_{\mathrm{even/odd}}(M)$ gives the number of closure phases $\Phi(\bm{m},\bm{n})$, i.e., essentially signs since the value of $|\Phi(\bm{m},\bm{n})|$ is known, and $M^2-3$ phases $\phi(\bm{m})$, since $\phi(0)=0$ and $\phi((1,0)^T)$, $\phi((0,1)^T)$ can be arbitrarily chosen, as they correspond to a global shift, i.e, they define the origin of the coordinate system [see Eq.~\eqref{eq:shift_dist}]. Let us consider the example of a camera with 9x9 pixels, i.e., $M=3$. In this case, we have $F^{(2)}_\mathrm{odd}(3) = 11$ non-trivial inequivalent equations, namely
\begin{align}
	\Phi((1,0)^T,(1,0)^T) &= \phi((2,0)^T) - 2 \phi((1,0)^T)\\
	\Phi((0,1)^T,(0,1)^T) &= \phi((0,2)^T) - 2 \phi((0,1)^T)\\
	\Phi((1,0)^T,(0,1)^T) &= \phi((1,1)^T) - \phi((1,0)^T) - \phi((0,1)^T)\\
	\Phi((1,0)^T,(1,1)^T) &= \phi((2,1)^T) - \phi((1,0)^T) - \phi((1,1)^T)\\
	\Phi((2,0)^T,(0,1)^T) &= \phi((2,1)^T) - \phi((2,0)^T) - \phi((0,1)^T)\\
	\Phi((0,1)^T,(1,1)^T) &= \phi((1,2)^T) - \phi((0,1)^T) - \phi((1,1)^T)\\
	\Phi((0,2)^T,(1,0)^T) &= \phi((1,2)^T) - \phi((0,2)^T) - \phi((1,0)^T)\\
	\Phi((1,0)^T,(1,2)^T) &= \phi((2,2)^T) - \phi((1,0)^T) - \phi((1,2)^T)\\
	\Phi((2,0)^T,(0,2)^T) &= \phi((2,2)^T) - \phi((2,0)^T) - \phi((0,2)^T)\\
	\Phi((2,1)^T,(0,1)^T) &= \phi((2,2)^T) - \phi((2,1)^T) - \phi((0,1)^T)\\
	\Phi((1,1)^T,(1,1)^T) &= \phi((2,2)^T) - 2 \phi((1,1)^T)\,,
\end{align}
but also $11$ unknown signs of the closure phases $\Phi(\bm{m},\bm{n})$ plus six unknown phases $\phi(\bm{m})$.\\
\noindent We now review the idea how to lift the sign ambiguity referring to Ref.~\cite{Peard23}. Let us consider again our 1D example of 9 pixels. Since $\phi(1)$ can be arbitrarily chosen, Eq.~\eqref{eq:11} gives two possible values for $\phi(2)$ due to the unknown sign of $\Phi(1,1)$. Further, Eq.~\eqref{eq:21} gives four possible values for $\phi(3)$ since $\phi(2)$ has two possible values and due to the unknown sign of $\Phi(2,1)$. Then, Eq.~\eqref{eq:31} gives eight possible values for $\phi(4)$. However, this equation is not the only equation including previous phases. Instead, Eq.~\eqref{eq:22} is an additional equation for $\phi(4)$ leading to four possible values. Since both equations have to give the same value for $\phi(4)$, the idea is to check for the intersection of the set containing the eight possible values coming from Eq.~\eqref{eq:31} and the set containing the four possible values coming from Eq.~\eqref{eq:22}. If there is only one equal number in both sets, we know that $\phi(4)$ has to be this number, which additionally determines also $\phi(2)$ and $\phi(3)$. We can then go on with Eqs.~\eqref{eq:41} and \eqref{eq:32} for $\phi(5)$ and check again the intersection. To find the remaining phases, we just go on with this procedure. In certain cases, this leads to a unique solution for the phase function of the structure factor, such that full information is obtained. However, there is an additional option to further reduce the ambiguity in the retrieval of the phase and to improve this algorithm, namely by considering in addition the fourth-order photon correlation function giving an additional restrictions.

\subsection{Fourth-order photon correlation function}
\noindent As we show in App.~\ref{sec:AppC}, the fourth-order photon correlation function leads to different \textit{additional} cosine terms of phases, i.e., it gives further restrictions to the possible phase values. Besides the usual closure phases as, for instance, $\phi(\bm{q}_1+\bm{q}_2)-\phi(\bm{q}_1)-\phi(\bm{q}_2)$, we find for example also terms like $\phi(\bm{q}_1+\bm{q}_2+\bm{q}_3)-\phi(\bm{q}_1)-\phi(\bm{q}_2)-\phi(\bm{q}_3)$ or $\phi(\bm{q}_1)-\phi(\bm{q}_3)+\phi(\bm{q}_2+\bm{q}_3)-\phi(\bm{q}_1+\bm{q}_2)$, which include four phase terms. To show how these additional terms can be beneficial, let us consider again the 1D example. Here, we will construct by hand a scenario where the fourth-order correlation function can resolve an ambiguity, which is not resolvable by the third-order correlation function. Even though this is a particular example, it shows how the fourth-order correlation function can be helpful if no unique intersection value is found for the reconstructed phases via the closure relation based on $g^{(3)}(\bm{k}_1,\bm{k}_2,\bm{k}_3)$. Let us assume that the first four closure phases are $\Phi(1,1)=\Phi(2,1)=\Phi(3,1)=\Phi(2,2)=\pm 1$. Without loss of generality, we can choose $\phi(1)=0$, such that $\phi(2)=\Phi(1,1)=\pm 1$. Let us neglect a possible point reflection of the configuration, such that we can keep the plus sign for simplicity. This leads to two possible values for $\phi(3)$, namely $\phi(3)\in \{2,0\}$ via Eq.~\eqref{eq:21}. Now, using Eq.~\eqref{eq:31} we find four possible values for $\phi(4)$, namely $\phi(4)\in \{3,1,1,-1\}$ with different combinations of $\Phi(3,1)$ and $\phi(3)$. However, via Eq.~\eqref{eq:22} we have an additional restriction for $\phi(4)$, which leads to the set of possible values given by $\phi(4)\in\{3,1\}$. As we find, the intersection of the two sets is not unique. However, considering the first non-equivalent value of $g^{(4)}$ we find (see App.~\ref{sec:AppC}:
\begin{widetext}
\begin{align}
    g^{(4)}(1,1,1) &= 1 - \frac{12}{N} + \frac{1}{N^2}(64+3|\tilde{S}(1)|^2+2|\tilde{S}(2)|^2+|\tilde{S}(3)|^2) + \frac{1}{N^3}\{-144-34|\tilde{S}(1)|^2-24|\tilde{S}(2)|^2-10|\tilde{S}(3)|^2\nonumber\\
    &+4|\tilde{S}(1)|^2|\tilde{S}(2)|\cos[\phi(2)-2\phi(1)]+4|\tilde{S}(1)||\tilde{S}(2)||\tilde{S}(3)|\cos[\phi(3)-\phi(2)-\phi(1)]\}\nonumber\\
    &+\frac{1}{N^4}\{|\tilde{S}(1)|^4 + |\tilde{S}(2)|^4 + 2|\tilde{S}(1)|^2|\tilde{S}(2)|^2 |\tilde{S}(1)|^2|\tilde{S}(3)|^2 + 96|\tilde{S}(1)|^2 + 68 |\tilde{S}(2)|^2 + 32 |\tilde{S}(3)|^2 + 4 |\tilde{S}(4)|^2\nonumber\\
    &-28 |\tilde{S}(1)|^2|\tilde{S}(2)|\cos[\phi(2)-2\phi(1)] -28 |\tilde{S}(1)||\tilde{S}(2)||\tilde{S}(3)|\cos[\phi(3)-\phi(2)-\phi(1)]\nonumber\\
    &-4 |\tilde{S}(2)|^2|\tilde{S}(4)| \cos[\phi(4)-2\phi(2)] - 4 |\tilde{S}(1)||\tilde{S}(3)||\tilde{S}(4)|\cos[\phi(4)-\phi(3)-\phi(1)]\nonumber\\
    &+2 |\tilde{S}(1)|^3|\tilde{S}(3)|\cos[\textcolor{magenta}{\phi(3)-3\phi(1)}] + 2 |\tilde{S}(1)||\tilde{S}(2)|^2|\tilde{S}(3)| \cos[\textcolor{magenta}{\phi(3)-2\phi(2)+\phi(1)}]\}
\end{align}
\end{widetext}
Thus, in the last line we find two new phase relations involving $\phi(3)$, namely $\phi(3)-3\phi(1)$ and $\phi(3)-2\phi(2)+\phi(1)$. If we come back to our explicit example, we would have $\phi(3)-3\phi(1)=\phi(3)\in \{2,0\}$ and $\phi(3)-2\phi(2)+\phi(1)=\phi(3)-2\phi(2)\in \{0,-2\}$. Consequently, since the cosines give different values for the different possible phase relations, it is possible to determine the correct value of $\phi(3)$ via a $g^{(4)}$ measurement, which then reduces the possible values for the other phases. Let us, for instance, assume that the measurement revealed that $\phi(3)=0$ is the correct value. Then, the possible values for $\phi(4)$ from Eq.~\eqref{eq:31} reduce to $\phi(4)\in\{1,-1\}$. Now, the intersection with the set $\{3,1\}$ is unique and we can immediately identify the value of $\phi(4)$ to be $1$. We emphasize again that the additional constraints given by the fourth-order correlation function $g^{(4)}(\bm{k}_1,\bm{k}_2,\bm{k}_3,\bm{k}_4)$ is always helpful if no unique intersection value is found for the reconstructed phases via the closure relation based on $g^{(3)}(\bm{k}_1,\bm{k}_2,\bm{k}_3)$.

\section{Conclusion}
\label{sec:Conclusion}
\noindent In conclusion, we have shown how higher-order photon correlation functions can be useful and beneficial to retrieve the phase of the structure factor $\tilde{S}(\bm{q})$, which is not accessible via CDI or IDI. To that aim, we have provided a complete quantum mechanical derivation of the photon correlation functions of order $3$ and $4$, displaying the closure phase and additional phase relations. We have demonstrated in particular that by using the fourth-order photon correlation function where new non-equivalent phase relations appear due to four-point correlations the phase retrieval process can be improved. Furthermore, we have shown by explicitly counting the number of equations and unknowns that even though the sign ambiguity of the closure phase equations can be lifted in certain cases, the closure phase equations are not unique. We believe that our findings provide new input in the route of using higher-order photon correlation functions for phase reconstruction.

\section*{Acknowledgements}
\noindent M.B. and J.v.Z. gratefully acknowledge funding and support by the International Max Planck Research School - Physics of Light. S.R. and J.v.Z. gratefully acknowledge funding and support by the Erlangen Graduate School in Advanced Optical Technologies (SAOT). This work was funded by the Deutsche Forschungsgemeinschaft (DFG, German Research Foundation) -- Project-ID 429529648 -- TRR 306 QuCoLiMa ("Quantum Cooperativity of Light and Matter'').\\
\\
\newpage

\appendix
\section{Quantum mechanical calculation of $g^{(2)}$}
\label{sec:AppA}
\noindent 
We consider the so-called normalized second-order photon correlation function
\begin{widetext}
\begin{align}
	g^{(2)}(\bm{k}_1,\bm{k}_2) &= \frac{G^{(2)}(\bm{k}_1,\bm{k}_2)}{G^{(1)}(\bm{k}_1,\bm{k}_1)G^{(1)}(\bm{k}_2,\bm{k}_2)} = \frac{1}{N^2}\braket{\op{E}^{(-)}(\bm{k}_1)\op{E}^{(-)}(\bm{k}_2)\op{E}^{(+)}(\bm{k}_2)\op{E}^{(+)}(\bm{k}_1)}\nonumber\\
	&= \frac{1}{N^2}\sum_{i,j,k,l=1}^N e^{i\bm{k}_1\bm{R}_i}e^{i\bm{k}_2\bm{R}_j}e^{-i\bm{k}_2\bm{R}_k}e^{-i\bm{k}_1\bm{R}_l}\braket{\op{S}_+^{(i)} \op{S}_+^{(j)} \op{S}_-^{(k)} \op{S}_-^{(l)}}\,.
	\label{eq:g2def}
\end{align}
\end{widetext}
Since the atoms are assumed to be initially in the excited state and emit incoherently, we only get a contribution if we have pairs of raising and lowering operators with the same index. In addition, since the atoms can emit at most one photon per pulsed excitation cycle, the case of all indices being equal gives zero. Therefore, we can write
\begin{align}
	\braket{\op{S}_+^{(i)} \op{S}_+^{(j)} \op{S}_-^{(k)} \op{S}_-^{(l)}}&=(1-\delta_{ij})(\delta_{ik}\delta_{jl}+\delta_{il}\delta_{jk})\,.\label{eq:G2Krondelta}
\end{align}
For the following calculations, it is convenient to describe the contributing terms by permutations of the corresponding symmetric permutation groups. With respect to the phase factors in Eq.~\eqref{eq:g2def} we can identify the index $i$ with "1" (index of $\bm{k}_1$), $j$ with "2", $k$ with "2", and $l$ with "1". Therefore, the Kronecker delta decomposition in Eq.~\eqref{eq:G2Krondelta} can be interpreted as the two permutation group elements of the symmetric permutation group $S_2$, where the Kronecker deltas correspond to the columns in Cauchy's two-line notation of a permutation. Explicitly, $\delta_{ik}$ corresponds to $1\rightarrow 2$, $\delta_{jl}$ to $2\rightarrow 1$, $\delta_{il}$ to $1 \rightarrow 1$, and $\delta_{jk}$ to $2 \rightarrow 2$. In Cauchy's two-line notation and cycle notation we can write
\begin{align}
	\delta_{ik}\delta_{jl}\; &\hat{=} \left(\begin{array}{cc}
		1 & 2 \\
		2 & 1
	\end{array}\right) = (1\;2)\,,\\
	\delta_{il}\delta_{jk}\; &\hat{=} \left(\begin{array}{cc}
		1 & 2 \\
		1 & 2
	\end{array}\right) = ()\,,
\end{align}
where the first line denotes the transposition of the two elements, whereas the second line is the identity. This notation will be especially useful when we consider the third- and fourth-order photon correlation functions.\\
As mentioned earlier, the case, in which all indices are equal is zero. In Eq.~\eqref{eq:G2Krondelta}, this is accounted for by including the factor $(1-\delta_{ij})$. Thus, we can write the second-order correlation function as
\begin{align}
	&g^{(2)}(\bm{k}_1,\bm{k}_2)\nonumber\\
 &= \frac{1}{N^2}\sum_{i,j=1}^N (1-\delta_{ij})\left(1+e^{-i(\bm{k}_2-\bm{k}_1)\bm{R}_i}e^{i(\bm{k}_2-\bm{k}_1)\bm{R}_j}\right)\,,
\end{align}
\normalsize
where we made use of Eq.~\eqref{eq:G2Krondelta}. Simplifying $g^{(2)}(\bm{k}_1,\bm{k}_2)$ leads us to
\begin{align}
	g^{(2)}(\bm{k}_1,\bm{k}_2)&=\frac{1}{N^2}\left[N^2-2N+\sum_{i,j=1}^N e^{-i\bm{q}\bm{R}_i}e^{i\bm{q}\bm{R}_j}\right]\nonumber\\
	&=1-\frac{2}{N}+\frac{1}{N^2}|\tilde{S}(\bm{q})|^2\,.
\end{align}
Note that for $\bm{k}_1=\bm{k}_2=\bm{k}$, i.e., a difference vector $\bm{q}=\bm{0}$, the correlation function reduces to
\begin{equation}
	g^{(2)}(\bm{k},\bm{k}) = 1-\frac{2}{N}+1=2\left(1-\frac{1}{N}\right)\,.
\end{equation}
Therefore, measuring the second-order photon correlation function for $\bm{q}=\bm{0}$ gives access to the number of atoms $N$. In addition, we find that measuring the second-order photon correlation function of the fluorescence light, i.e., utilizing incoherent diffractive imaging, gives access to the magnitude of the CDC.

\section{Quantum mechanical calculation of $g^{(3)}$}
\label{sec:AppB}
\noindent Next, we calculate the normalized third-order photon correlation function for $N$ initially excited atoms, defined as
\begin{widetext}
\begin{align}
	g^{(3)}(\bm{k}_1,\bm{k}_2,\bm{k}_3)&=\frac{G^{(3)}(\bm{k}_1,\bm{k}_2,\bm{k}_3)}{G^{(1)}(\bm{k}_1,\bm{k}_1)G^{(1)}(\bm{k}_2,\bm{k}_2)G^{(1)}(\bm{k}_3,\bm{k}_3)}=\frac{1}{N^3}\braket{\op{E}^{(-)}(\bm{k}_1)\op{E}^{(-)}(\bm{k}_2)\op{E}^{(-)}(\bm{k}_3)\op{E}^{(+)}(\bm{k}_3)\op{E}^{(+)}(\bm{k}_2)\op{E}^{(+)}(\bm{k}_1)}\nonumber\\
	&=\frac{1}{N^3}\sum_{i,j,k,l,m,n=1}^N e^{i\bm{k}_1\bm{R}_i}e^{i\bm{k}_2\bm{R}_j}e^{i\bm{k}_3\bm{R}_k}e^{-i\bm{k}_3\bm{R}_l}e^{-i\bm{k}_2\bm{R}_m}e^{-i\bm{k}_1\bm{R}_n}\braket{\op{S}_+^{(i)} \op{S}_+^{(j)} \op{S}_+^{(k)} \op{S}_-^{(l)} \op{S}_-^{(m)} \op{S}_-^{(n)}}\,.
\end{align}
\end{widetext}
As before, we only get a contribution if we have pairs of raising and lowering operators, which have the same index, whereas the expectation value vanishes in the case that two pairs have the same index. In particular, we can write
\begin{widetext}
\begin{equation}
\braket{\op{S}_+^{(i)} \op{S}_+^{(j)} \op{S}_+^{(k)} \op{S}_-^{(l)} \op{S}_-^{(m)} \op{S}_-^{(n)}}=\delta_{il}\delta_{jm}\delta_{kn}+\delta_{il}\delta_{jn}\delta_{km}+\delta_{im}\delta_{jl}\delta_{kn}+\delta_{im}\delta_{jn}\delta_{kl}+\delta_{in}\delta_{jl}\delta_{km}+\delta_{in}\delta_{jm}\delta_{kl}
\end{equation}
\end{widetext}
for $i\neq j,i\neq k,j\neq k$ pairwise different. We can identify the different delta contributions with permutation elements of the symmetric permutation group $S_3$. In Cauchy's two-line notation and in cycle notation we find
\begin{align}
	\delta_{il}\delta_{jm}\delta_{kn}\; &\hat{=} \left(\begin{array}{ccc}
		1 & 2 & 3 \\
		3 & 2 & 1
	\end{array}\right) = (1\;3)\,,\nonumber\\
	\delta_{il}\delta_{jn}\delta_{km}\; &\hat{=} \left(\begin{array}{ccc}
		1 & 2 & 3 \\
		3 & 1 & 2
	\end{array}\right) = (1\;3\;2)\,,\nonumber\\
	\delta_{im}\delta_{jl}\delta_{kn}\; &\hat{=} \left(\begin{array}{ccc}
		1 & 2 & 3 \\
		2 & 3 & 1
	\end{array}\right) = (1\;2\;3)\,,\nonumber\\
	\delta_{im}\delta_{jn}\delta_{kl}\; &\hat{=} \left(\begin{array}{ccc}
		1 & 2 & 3 \\
		2 & 1 & 3
	\end{array}\right) = (1\;2)\,,\nonumber\\
	\delta_{in}\delta_{jl}\delta_{km}\; &\hat{=} \left(\begin{array}{ccc}
		1 & 2 & 3 \\
		1 & 3 & 2
	\end{array}\right) = (2\;3)\,,\nonumber\\
	\delta_{in}\delta_{jm}\delta_{kl}\; &\hat{=} \left(\begin{array}{ccc}
		1 & 2 & 3 \\
		1 & 2 & 3
	\end{array}\right) = ()\,.\nonumber
\end{align}
Therefore, we only have to calculate the sum for three of the summands corresponding to the identity, one of the transpositions, and one of the 3-cycles. The remaining three summands are then easily deduced by index replacements. We note again that we have to restrict the summation to all $i,j,k$ being mutually different, which is accounted for by multiplication with the factor $(1-\delta_{ij})(1-\delta_{ik})(1-\delta_{jk})$. Note that we can write this product also as
\begin{equation}
	(1-\delta_{ij})(1-\delta_{ik})(1-\delta_{jk})=1+\delta_{ij} \delta_{ik}+\delta_{ij} \delta_{jk}-\delta_{ij}-\delta_{ik}-\delta_{jk}\,.
\end{equation}
We first consider the identity term
\begin{equation}
	\sum_{i,j,k=1}^N (1+\delta_{ij} \delta_{ik}+\delta_{ij} \delta_{jk}-\delta_{ij}-\delta_{ik}-\delta_{jk})=N^3+2N-3N^2\,.
\end{equation}
Second, we choose the transposition term $(1\;2)$, which gives
\begin{align}
	\sum_{i,j,k=1}^N (1+\delta_{ij} \delta_{ik}+\delta_{ij} \delta_{jk}-\delta_{ij}-\delta_{ik}-\delta_{jk})e^{-i\bm{q}_1\bm{R}_i}e^{i\bm{q}_1\bm{R}_j}\nonumber\\
	=2N-N^2+(N-2)|\tilde{S}(\bm{q}_1)|^2\,,
\end{align}
where we defined $\bm{q}_1=\bm{k}_2-\bm{k}_1$. Last, we need one of the 3-cycle terms. We choose $(1\;3\;2)$, which gives
\begin{widetext}
\begin{align}
	&\sum_{i,j,k=1}^N (1+\delta_{ij} \delta_{ik}+\delta_{ij} \delta_{jk}-\delta_{ij}-\delta_{ik}-\delta_{jk})e^{-i\bm{q}_3\bm{R}_i}e^{i\bm{q}_2\bm{R}_k}e^{i\bm{q}_1\bm{R}_j}\nonumber\\
	&=\tilde{S}^*(\bm{q}_1)\tilde{S}^*(\bm{q}_2)\tilde{S}_r(\bm{q}_3)-\tilde{S}^*(\bm{q}_1+\bm{q}_2)\tilde{S}_r(\bm{q}_3)-\tilde{S}^*(\bm{q}_2)\tilde{S}_r(\bm{q}_3-\bm{q}_1)-\tilde{S}^*(\bm{q}_1)\tilde{S}_r(\bm{q}_3-\bm{q}_2)+2\tilde{S}_r(\bm{q}_3-\bm{q}_1-\bm{q}_2)\nonumber\\
	&=2N-|\tilde{S}(\bm{q}_1)|^2-|\tilde{S}(\bm{q}_2)|^2-|\tilde{S}(\bm{q}_1+\bm{q}_2)|^2+\tilde{S}^*(\bm{q}_1)\tilde{S}^*(\bm{q}_2)\tilde{S}_r(\bm{q}_1+\bm{q}_2)\,,
\end{align}
\end{widetext}
where $\bm{q}_2=\bm{k}_3-\bm{k}_2$ and $\bm{q}_3=\bm{k}_3-\bm{k}_1=\bm{k}_3-\bm{k}_2+\bm{k}_2-\bm{k}_1=\bm{q}_1+\bm{q}_2$. Accounting for all six permutations $\textcolor{Dark Pastel Red}{()}$, $\textcolor{Carrot Orange}{(1\;2)}$, $\textcolor{Carrot Orange}{(2\;3)}$, $\textcolor{Carrot Orange}{(1\;3)}$, $\textcolor{Dark Pastel Purple}{(1\;3\;2)}$, and $\textcolor{Dark Pastel Purple}{(1\;2\;3)}$, the third-order photon correlation function calculates to
\begin{widetext}
\begin{align}
	g^{(3)}(\bm{k}_1,\bm{k}_2,\bm{k}_3)&=\frac{1}{N^3}\left[\textcolor{Dark Pastel Red}{N^3+2N-3N^2}+\textcolor{Carrot Orange}{2N-N^2+(N-2)|\tilde{S}(\bm{q}_1)|^2}\nonumber\right.\\
	&\left.+\textcolor{Carrot Orange}{2N-N^2+(N-2)|\tilde{S}(\bm{q}_2)|^2}+\textcolor{Carrot Orange}{2N-N^2+(N-2)|\tilde{S}(\bm{q}_1+\bm{q}_2)|^2}\right.\nonumber\\
	&\left.+\textcolor{Dark Pastel Purple}{2N-|\tilde{S}(\bm{q}_1)|^2-|\tilde{S}(\bm{q}_2)|^2-|\tilde{S}(\bm{q}_1+\bm{q}_2)|^2+\tilde{S}^*(\bm{q}_1)\tilde{S}^*(\bm{q}_2)\tilde{S}_r(\bm{q}_1+\bm{q}_2)}\right.\nonumber\\
	&\left.+\textcolor{Dark Pastel Purple}{2N-|\tilde{S}(\bm{q}_1)|^2-|\tilde{S}(\bm{q}_2)|^2-|\tilde{S}(\bm{q}_1+\bm{q}_2)|^2+\tilde{S}_r(\bm{q}_1)\tilde{S}_r(\bm{q}_2)\tilde{S}^*(\bm{q}_1+\bm{q}_2)}\right]\nonumber\\
	&=1-\frac{6}{N}+\frac{12}{N^2}+\frac{N-4}{N^3}[|\tilde{S}(\bm{q}_1)|^2+|\tilde{S}(\bm{q}_2)|^2+|\tilde{S}(\bm{q}_1+\bm{q}_2)|^2]\nonumber\\
	&+ \frac{2}{N^3}|\tilde{S}(\bm{q}_1)||\tilde{S}(\bm{q}_2)||\tilde{S}(\bm{q}_1+\bm{q}_2)| \cos[\phi(\bm{q}_1+\bm{q}_2)-\phi(\bm{q}_1)-\phi(\bm{q}_2)]\,,
	\label{eq:g3}
\end{align}
\end{widetext}
where we have written the structure factor in terms of its magnitude and phase as
\begin{equation}
	\tilde{S}(\bm{q})=|\tilde{S}(\bm{q})|e^{i\phi(\bm{q})}\,,
\end{equation}
and also used 
\begin{equation}
	|\tilde{S}(-\bm{q})|=|\tilde{S}(\bm{q})|\,,\quad \phi(-\bm{q})=-\phi(\bm{q})\,,
\end{equation}
which holds for an arbitrary real-valued atomic distribution $S(\bm{r})$.

\section{Quantum mechanical calculation of $g^{(4)}$}
\label{sec:AppC}
\noindent We consider the normalized fourth-order photon correlation function defined as
\begin{widetext}
\begin{align}
	&g^{(4)}(\bm{k}_1,\bm{k}_2,\bm{k}_3,\bm{k}_4)=\frac{G^{(4)}(\bm{k}_1,\bm{k}_2,\bm{k}_3,\bm{k}_4)}{G^{(1)}(\bm{k}_1,\bm{k}_1)G^{(1)}(\bm{k}_2,\bm{k}_2)G^{(1)}(\bm{k}_3,\bm{k}_3)G^{(1)}(\bm{k}_4,\bm{k}_4)}\nonumber\\
	&=\frac{1}{N^4}\braket{\op{E}^{(-)}(\bm{k}_1)\op{E}^{(-)}(\bm{k}_2)\op{E}^{(-)}(\bm{k}_3)\op{E}^{(-)}(\bm{k}_4)\op{E}^{(+)}(\bm{k}_4)\op{E}^{(+)}(\bm{k}_3)\op{E}^{(+)}(\bm{k}_2)\op{E}^{(+)}(\bm{k}_1)}\nonumber\\
	&=\frac{1}{N^4}\sum_{i,j,k,l,m,n,o,p=1}^N e^{i\bm{k}_1\bm{R}_i}e^{i\bm{k}_2\bm{R}_j}e^{i\bm{k}_3\bm{R}_k}e^{i\bm{k}_4\bm{R}_l}e^{-i\bm{k}_4\bm{R}_m}e^{-i\bm{k}_3\bm{R}_n}e^{-i\bm{k}_2\bm{R}_o}e^{-i\bm{k}_1\bm{R}_p}\braket{\op{S}_+^{(i)} \op{S}_+^{(j)} \op{S}_+^{(k)} \op{S}_+^{(l)} \op{S}_-^{(m)} \op{S}_-^{(n)} \op{S}_-^{(o)} \op{S}_-^{(p)}}\,.
\end{align}
\end{widetext}
Again, the expectation value of the normally ordered product of raising and lowering operators is only non-zero if we have pairs of raising and lowering operators with the same index. This gives $4!=24$ different contributions, which correspond to the $24$ permutation elements of the symmetric permutation group $S_4$. The group consists of one \textcolor{Dark Pastel Red}{identity} permutation, six \textcolor{Carrot Orange}{transpositions}, three \textcolor{Minion Yellow}{double transpositions}, eight \textcolor{Dark Pastel Green}{3-cycles} and six \textcolor{Silver Lake Blue}{4-cycles}, which are listed in Tab.~\ref{tab:S4} in cycle notation.
\begin{table*}[]
	\centering
	\begin{tabular}{c|c}
		permutation type & elements in cycle notation\\
		\hline
		identity & $()$\\
		transposition & $(1\;2)$, $(1\;3)$, $(1\;4)$, $(2\;3)$, $(2\;4)$, $(3\;4)$\\
		double transposition & $(1\;2)(3\;4)$, $(1\;3)(2\;4)$, $(1\;4)(2\;3)$\\
		3-cycle & $(1\;2\;3)$, $(1\;3\;2)$, $(2\;3\;4)$, $(2\;4\;3)$, $(3\;4\;1)$, $(3\;1\;4)$, $(4\;1\;2)$, $(4\;2\;1)$\\
		4-cycle & $(1\;2\;3\;4)$, $(1\;2\;4\;3)$, $(1\;3\;2\;4)$, $(1\;3\;4\;2)$, $(1\;4\;2\;3)$, $(1\;4\;3\;2)$
	\end{tabular}
	\caption{Elements of the symmetric permutation group $S_4$.}
	\label{tab:S4}
\end{table*}
Therefore, we find that we only have to calculate five different terms corresponding to the five different permutation types. In addition, the normally ordered expectation value gives zero if two creation and annihilation operator pairs have the same index. Thus, the remaining indices $i,j,k,l$ need to be mutually different. This is accounted for by multiplication with
\begin{widetext}
\begin{align}
	&\text{del}(i,j,k,l)\coloneqq(1-\delta_{ij})(1-\delta_{ik})(1-\delta_{il})(1-\delta_{jk})(1-\delta_{jl})(1-\delta_{kl})\nonumber\\
	&=1-\delta _{ij} \delta _{ik} \delta _{il}-\delta _{ij} \delta _{il} \delta _{jk}-\delta _{ij} \delta _{ik} \delta _{jl}-\delta _{ij} \delta _{jk} \delta _{jl}-\delta _{ij} \delta
	_{ik} \delta _{kl}-\delta _{ij} \delta _{jk} \delta _{kl}+\delta _{ij} \delta _{kl}+\delta _{il} \delta _{jk}+\delta _{ik} \delta _{jl}\nonumber\\
	&+\delta _{ij} \delta _{ik}+\delta _{ij}
	\delta _{jk}+\delta _{ij} \delta _{il}+\delta _{ij} \delta _{jl}+\delta _{ik} \delta _{il}+\delta _{ik} \delta _{kl}+\delta _{jk} \delta
	_{jl}+\delta _{jk} \delta _{kl}-\delta _{ij}-\delta _{ik}-\delta _{il}-\delta _{jk}-\delta _{jl}-\delta _{kl}\,.
\end{align}
Now, we evaluate the five different contribution types.\\
The \textcolor{Dark Pastel Red}{identity} type gives
\begin{align}
	\textcolor{Dark Pastel Red}{p_1}=\sum_{i,j,k,l=1}^N \text{del}(i,j,k,l) = N^4 - 6 N^3 + 11 N^2 -6 N\,.
\end{align}
The \textcolor{Carrot Orange}{transposition} type yields
\begin{align}
	\textcolor{Carrot Orange}{p_2(\bm{q})}&=\sum_{i,j,k,l=1}^N \text{del}(i,j,k,l)e^{-i\bm{q}(\bm{R}_k-\bm{R}_l)}=N^2 \tilde{S}^*(\bm{q}) \tilde{S}(\bm{q})-N^3+5 N^2-5 N \tilde{S}^*(\bm{q}) \tilde{S}(\bm{q})-6 N+6 \tilde{S}^*(\bm{q}) \tilde{S}(\bm{q})\nonumber\\
	&=-N^3+5N^2-6N+(N^2-5N+6)|\tilde{S}(\bm{q})|^2\,.
\end{align}
The \textcolor{Minion Yellow}{double transposition} type is given by
\begin{align}
	\textcolor{Minion Yellow}{p_3(\bm{q}_1,\bm{q}_2)}&=\sum_{i,j,k,l=1}^N \text{del}(i,j,k,l)e^{-i \bm{q}_1 (\bm{R}_i-\bm{R}_j)-i\bm{q}_2 (\bm{R}_k-\bm{R}_l)}\nonumber\\
	&=N^2-N \tilde{S}^*(\bm{q}_1) \tilde{S}(\bm{q}_1)-N \tilde{S}^*(\bm{q}_2) \tilde{S}(\bm{q}_2)-6 N+\tilde{S}^*(\bm{q}_1) \tilde{S}(\bm{q}_1) \tilde{S}^*(\bm{q}_2) \tilde{S}(\bm{q}_2)-\tilde{S}^*(\bm{q}_1) \tilde{S}(\bm{q}_2)\tilde{S}^*(\bm{q}_2-\bm{q}_1)\nonumber\\
	&-\tilde{S}(\bm{q}_1) \tilde{S}^*(\bm{q}_2) \tilde{S}(\bm{q}_2-\bm{q}_1)-\tilde{S}^*(\bm{q}_1) \tilde{S}^*(\bm{q}_2) \tilde{S}(\bm{q}_1+\bm{q}_2)-\tilde{S}(\bm{q}_1)\tilde{S}(\bm{q}_2) \tilde{S}^*(\bm{q}_1+\bm{q}_2)+\tilde{S}^*(\bm{q}_2-\bm{q}_1) \tilde{S}(\bm{q}_2-\bm{q}_1)\nonumber\\
	&+\tilde{S}^*(\bm{q}_1+\bm{q}_2) \tilde{S}(\bm{q}_1+\bm{q}_2)+4 \tilde{S}^*(\bm{q}_1)
	\tilde{S}(\bm{q}_1)+4 \tilde{S}^*(\bm{q}_2) \tilde{S}(\bm{q}_2)\nonumber\\
	&=N^2-6N-(N-4)[|\tilde{S}(\bm{q}_1)|^2+|\tilde{S}(\bm{q}_2)|^2]+|\tilde{S}(\bm{q}_2-\bm{q}_1)|^2+|\tilde{S}(\bm{q}_1+\bm{q}_2)|^2\nonumber\\
	&+|\tilde{S}(\bm{q}_1)|^2|\tilde{S}(\bm{q}_2)|^2-2|\tilde{S}(\bm{q}_1)||\tilde{S}(\bm{q}_2)||\tilde{S}(\bm{q}_2-\bm{q}_1)|\cos[\phi(\bm{q}_1)-\phi(\bm{q}_2)+\phi(\bm{q}_2-\bm{q}_1)]\nonumber\\
	&-2|\tilde{S}(\bm{q}_1)||\tilde{S}(\bm{q}_2)||\tilde{S}(\bm{q}_1+\bm{q}_2)|\cos[\phi(\bm{q}_1)+\phi(\bm{q}_2)-\phi(\bm{q}_1+\bm{q}_2)]\,.
\end{align}
The \textcolor{Dark Pastel Green}{3-cycle} type reads
\begin{align}
	\textcolor{Dark Pastel Green}{p_4(\bm{q}_1,\bm{q}_2,\bm{q}_3)}&=\sum_{i,j,k,l=1}^N \text{del}(i,j,k,l)e^{-i \bm{q}_3 \bm{R}_j+i\bm{q}_1 \bm{R}_k+i\bm{q}_2 \bm{R}_l}\nonumber\\
	&=N \tilde{S}^*(\bm{q}_1) \tilde{S}^*(\bm{q}_2) \tilde{S}(\bm{q}_3)-N \tilde{S}^*(\bm{q}_2) \tilde{S}(\bm{q}_3-\bm{q}_1)-N \tilde{S}^*(\bm{q}_1) \tilde{S}(\bm{q}_3-\bm{q}_2)-N \tilde{S}(\bm{q}_3)
	\tilde{S}^*(\bm{q}_1+\bm{q}_2)\nonumber\\
	&+2 N \tilde{S}(\bm{q}_3-\bm{q}_1-\bm{q}_2)-3 \tilde{S}^*(\bm{q}_1) \tilde{S}^*(\bm{q}_2) \tilde{S}(\bm{q}_3)+3 \tilde{S}^*(\bm{q}_2) \tilde{S}(\bm{q}_3-\bm{q}_1)+3 \tilde{S}^*(\bm{q}_1)\tilde{S}(\bm{q}_3-\bm{q}_2)\nonumber\\
	&+3 \tilde{S}(\bm{q}_3) \tilde{S}^*(\bm{q}_1+\bm{q}_2)-6 \tilde{S}(\bm{q}_3-\bm{q}_1-\bm{q}_2)\nonumber\\
	&=(N-3)[\tilde{S}^*(\bm{q}_1) \tilde{S}^*(\bm{q}_2) \tilde{S}(\bm{q}_3)-\tilde{S}^*(\bm{q}_2) \tilde{S}(\bm{q}_3-\bm{q}_1)-\tilde{S}^*(\bm{q}_1) \tilde{S}(\bm{q}_3-\bm{q}_2)\nonumber\\
	&-\tilde{S}(\bm{q}_3)
	\tilde{S}^*(\bm{q}_1+\bm{q}_2)+2\tilde{S}(\bm{q}_3-\bm{q}_1-\bm{q}_2)]\,.
\end{align}
The \textcolor{Silver Lake Blue}{4-cycle} type calculates to
\begin{align}
	\textcolor{Silver Lake Blue}{p_5(\bm{q}_1,\bm{q}_2,\bm{q}_3,\bm{q}_4)}&=\sum_{i,j,k,l=1}^N \text{del}(i,j,k,l)e^{-i \bm{q}_4 \bm{R}_i+i\bm{q}_1 \bm{R}_j+i\bm{q}_2 \bm{R}_k+i\bm{q}_3 \bm{R}_l}\nonumber\\
	&=\tilde{S}^*(\bm{q}_1) \tilde{S}^*(\bm{q}_2) \tilde{S}^*(\bm{q}_3) \tilde{S}(\bm{q}_4)-\tilde{S}^*(\bm{q}_3) \tilde{S}(\bm{q}_4) \tilde{S}^*(\bm{q}_1+\bm{q}_2)-\tilde{S}^*(\bm{q}_2) \tilde{S}(\bm{q}_4)\tilde{S}^*(\bm{q}_1+\bm{q}_3)\nonumber\\
	&-\tilde{S}^*(\bm{q}_1) \tilde{S}(\bm{q}_4) \tilde{S}^*(\bm{q}_2+\bm{q}_3)-\tilde{S}^*(\bm{q}_2) \tilde{S}^*(\bm{q}_3) \tilde{S}(\bm{q}_4-\bm{q}_1)-\tilde{S}^*(\bm{q}_1) \tilde{S}^*(\bm{q}_3)
	\tilde{S}(\bm{q}_4-\bm{q}_2)\nonumber\\
	&-\tilde{S}^*(\bm{q}_1) \tilde{S}^*(\bm{q}_2) \tilde{S}(\bm{q}_4-\bm{q}_3)+2 \tilde{S}^*(\bm{q}_3) \tilde{S}(\bm{q}_4-\bm{q}_1-\bm{q}_2)+2 \tilde{S}^*(\bm{q}_2)\tilde{S}(\bm{q}_4-\bm{q}_1-\bm{q}_3)\nonumber\\
	&+2 \tilde{S}^*(\bm{q}_1) \tilde{S}(\bm{q}_4-\bm{q}_2-\bm{q}_3)+2 \tilde{S}(\bm{q}_4)
	\tilde{S}^*(\bm{q}_1+\bm{q}_2+\bm{q}_3)+\tilde{S}(\bm{q}_4-\bm{q}_1) \tilde{S}^*(\bm{q}_2+\bm{q}_3)\nonumber\\
	&+\tilde{S}^*(\bm{q}_1+\bm{q}_3)
	\tilde{S}(\bm{q}_4-\bm{q}_2)+\tilde{S}^*(\bm{q}_1+\bm{q}_2) \tilde{S}(\bm{q}_4-\bm{q}_3)-6 \tilde{S}(\bm{q}_4-\bm{q}_1-\bm{q}_2-\bm{q}_3)\,.
\end{align}
Having the five different contribution types evaluated, we define the difference vectors $\bm{q}_1=\bm{k}_2-\bm{k}_1$, $\bm{q}_2=\bm{k}_3-\bm{k}_2$, $\bm{q}_3=\bm{k}_4-\bm{k}_3$, $\bm{q}_4=\bm{k}_3-\bm{k}_1=\bm{q}_1+\bm{q}_2$, $\bm{q}_5=\bm{k}_4-\bm{k}_1=\bm{q}_1+\bm{q}_2+\bm{q}_3$, $\bm{q}_6=\bm{k}_4-\bm{k}_2=\bm{q}_2+\bm{q}_3$. Now, the normalized fourth-order photon correlation function is simply the sum over the $24$ terms corresponding to the different permutations of $S_4$, which can be calculated via the functions $p_1$ to $p_5$ by properly plugging in the correct difference vectors for the different permutations, i.e.,
\begin{align}
	&N^4 g^{(4)}(\bm{k}_1,\bm{k}_2,\bm{k}_3,\bm{k}_4)=\textcolor{Dark Pastel Red}{p_1}\textcolor{Carrot Orange}{+p_2(\bm{q}_1)+p_2(\bm{q}_2)+p_2(\bm{q}_3)+p_2(\bm{q}_4)+p_2(\bm{q}_5)+p_2(\bm{q}_6)}\nonumber\\
	&\textcolor{Minion Yellow}{+p_3(\bm{q}_1,\bm{q}_3)+p_3(\bm{q}_2,\bm{q}_5)+p_3(\bm{q}_4,\bm{q}_6)}\textcolor{Dark Pastel Green}{+p_4(\bm{q}_1,\bm{q}_2,\bm{q}_4)+p_4(-\bm{q}_2,\bm{q}_4,\bm{q}_1)+p_4(\bm{q}_1,
		\bm{q}_6,\bm{q}_5)}\nonumber\\
	&\textcolor{Dark Pastel Green}{+p_4(-\bm{q}_6,\bm{q}_5,\bm{q}_1)+p_4(\bm{q}_2,\bm{q}_3,\bm{q}_6)+p_4(-\bm{q}_3,\bm{q}_6,\bm{q}_2)+p_4(-\bm{q}_3,\bm{q}_5,\bm{q}_4)+p_4(\bm{q}_4,\bm{q}_3,\bm{q}_5)}\nonumber\\
	&\textcolor{Silver Lake Blue}{+p_5(\bm{q}_1,\bm{q}_2,\bm{q}_3,\bm{q}_5)+p_5(-\bm{q}_2,-\bm{q}_3,\bm{q}_5,\bm{q}_1)+p_5(\bm{q}_1,-\bm{q}_3,\bm{q}_6,\bm{q}_4)}\nonumber\\
	&\textcolor{Silver Lake Blue}{+p_5(-\bm{q}_6,\bm{q}_4,\bm{q}_3,\bm{q}_1)+p_5(-\bm{q}_2,\bm{q}_4,\bm{q}_6,\bm{q}_5)+p_5(-\bm{q}_6,\bm{q}_2,\bm{q}_5,\bm{q}_4)}\,.
\end{align}
Thereby, the individual contributions are given by
\begin{align}
	\textcolor{Dark Pastel Red}{p_1}=N^4-6 N^3+11 N^2-6 N\,,
\end{align}
\begin{align}
	&\textcolor{Carrot Orange}{p_2(\bm{q}_1)+p_2(\bm{q}_2)+p_2(\bm{q}_3)+p_2(\bm{q}_4)+p_2(\bm{q}_5)+p_2(\bm{q}_6)}=\nonumber\\
	& -6 N^3+30 N^2-36 N+(N^2-5N+6) [|\tilde{S}(\bm{q}_1+\bm{q}_2+\bm{q}_3)|^2\nonumber\\
	&+|\tilde{S}(\bm{q}_1+\bm{q}_2)|^2+|\tilde{S}(\bm{q}_1)|^2+|\tilde{S}(\bm{q}_2+\bm{q}_3)|^2+|\tilde{S}(\bm{q}_2)|^2+|\tilde{S}(\bm{q}_3)|^2]\,,
\end{align}
\begin{align}
	&\textcolor{Minion Yellow}{p_3(\bm{q}_1,\bm{q}_3)+p_3(\bm{q}_2,\bm{q}_5)+p_3(\bm{q}_4,\bm{q}_6)}=3 N^2-18 N\nonumber\\
	&+|\tilde{S}(\bm{q}_1)|^2 |\tilde{S}(\bm{q}_3)|^2+|\tilde{S}(\bm{q}_1+\bm{q}_2)|^2 |\tilde{S}(\bm{q}_2+\bm{q}_3)|^2+|\tilde{S}(\bm{q}_2)|^2
		|\tilde{S}(\bm{q}_1+\bm{q}_2+\bm{q}_3)|^2\nonumber\\
	&-(N-4)[|\tilde{S}(\bm{q}_1+\bm{q}_2+\bm{q}_3)|^2+ |\tilde{S}(\bm{q}_1+\bm{q}_2)|^2+ |\tilde{S}(\bm{q}_1)|^2+ |\tilde{S}(\bm{q}_2+\bm{q}_3)|^2+ |\tilde{S}(\bm{q}_2)|^2+
		|\tilde{S}(\bm{q}_3)|^2]\nonumber\\
	&+2[ |\tilde{S}(\bm{q}_1+2 \bm{q}_2+\bm{q}_3)|^2+ |\tilde{S}(\bm{q}_3-\bm{q}_1)|^2+
		|\tilde{S}(\bm{q}_1+\bm{q}_3)|^2]\nonumber\\
	&-2\lbrace |\tilde{S}(\bm{q}_1)|
		|\tilde{S}(\bm{q}_3)| |\tilde{S}(\bm{q}_3-\bm{q}_1)| \cos [\phi (\bm{q}_1)-\phi (\bm{q}_3)+\phi (\bm{q}_3-\bm{q}_1)]\nonumber\\
	&+ |\tilde{S}(\bm{q}_1)| |\tilde{S}(\bm{q}_3)|
		|\tilde{S}(\bm{q}_1+\bm{q}_3)| \cos [\phi (\bm{q}_1)+\phi (\bm{q}_3)-\phi (\bm{q}_1+\bm{q}_3)]\nonumber\\
	&+ |\tilde{S}(\bm{q}_1+\bm{q}_2)| |\tilde{S}(\bm{q}_3-\bm{q}_1)| |\tilde{S}(\bm{q}_2+\bm{q}_3)| \cos [\phi (\bm{q}_1+\bm{q}_2)+\phi (\bm{q}_3-\bm{q}_1)-\phi
		(\bm{q}_2+\bm{q}_3)]\nonumber\\
	&+ |\tilde{S}(\bm{q}_2)| |\tilde{S}(\bm{q}_1+\bm{q}_3)| |\tilde{S}(\bm{q}_1+\bm{q}_2+\bm{q}_3)| \cos [\phi
		(\bm{q}_2)+\phi (\bm{q}_1+\bm{q}_3)-\phi (\bm{q}_1+\bm{q}_2+\bm{q}_3)]\nonumber\\
	&+ |\tilde{S}(\bm{q}_1+\bm{q}_2)| |\tilde{S}(\bm{q}_2+\bm{q}_3)| |\tilde{S}(\bm{q}_1+2 \bm{q}_2+\bm{q}_3)| \cos [\phi
		(\bm{q}_1+\bm{q}_2)+\phi (\bm{q}_2+\bm{q}_3)-\phi (\bm{q}_1+2 \bm{q}_2+\bm{q}_3)]\nonumber\\
	&+ |\tilde{S}(\bm{q}_2)| |\tilde{S}(\bm{q}_1+\bm{q}_2+\bm{q}_3)| |\tilde{S}(\bm{q}_1+2 \bm{q}_2+\bm{q}_3)| \cos [\phi (\bm{q}_2)+\phi
		(\bm{q}_1+\bm{q}_2+\bm{q}_3)-\phi (\bm{q}_1+2 \bm{q}_2+\bm{q}_3)]\rbrace\,,
\end{align}
\begin{align}
	&\textcolor{Dark Pastel Green}{p_4(\bm{q}_1,\bm{q}_2,\bm{q}_4)+p_4(-\bm{q}_2,\bm{q}_4,\bm{q}_1)+p_4(\bm{q}_1,
		\bm{q}_6,\bm{q}_5)+p_4(-\bm{q}_6,\bm{q}_5,\bm{q}_1)+p_4(\bm{q}_2,\bm{q}_3,\bm{q}_6)}\nonumber\\
	&\textcolor{Dark Pastel Green}{+p_4(-\bm{q}_3,\bm{q}_6,\bm{q}_2)+p_4(-\bm{q}_3,\bm{q}_5,\bm{q}_4)+p_4(\bm{q}_4,\bm{q}_3,\bm{q}_5)}=16 N^2-48 N\nonumber\\
	&-4(N-3)[|\tilde{S}(\bm{q}_1+\bm{q}_2+\bm{q}_3)|^2+|\tilde{S}(\bm{q}_1+\bm{q}_2)|^2+|\tilde{S}(\bm{q}_1)|^2+|\tilde{S}(\bm{q}_2+\bm{q}_3)|^2+|\tilde{S}(\bm{q}_2)|^2+|\tilde{S}(\bm{q}_3)|^2]\nonumber\\
	&+2 (N-3)\lbrace |\tilde{S}(\bm{q}_2)| |\tilde{S}(\bm{q}_3)| |\tilde{S}(\bm{q}_2+\bm{q}_3)| \cos [\phi (\bm{q}_2)+\phi
		(\bm{q}_3)-\phi (\bm{q}_2+\bm{q}_3)]\nonumber\\
	&+ |\tilde{S}(\bm{q}_1)| |\tilde{S}(\bm{q}_2)| |\tilde{S}(\bm{q}_1+\bm{q}_2)| \cos [\phi (\bm{q}_1)+\phi (\bm{q}_2)-\phi (\bm{q}_1+\bm{q}_2)]\nonumber\\
	&+ |\tilde{S}(\bm{q}_3)| |\tilde{S}(\bm{q}_1+\bm{q}_2)| |\tilde{S}(\bm{q}_1+\bm{q}_2+\bm{q}_3)| \cos [\phi (\bm{q}_3)+\phi (\bm{q}_1+\bm{q}_2)-\phi (\bm{q}_1+\bm{q}_2+\bm{q}_3)]\nonumber\\
	&+ |\tilde{S}(\bm{q}_1)| |\tilde{S}(\bm{q}_2+\bm{q}_3)| |\tilde{S}(\bm{q}_1+\bm{q}_2+\bm{q}_3)| \cos [\phi (\bm{q}_1)+\phi (\bm{q}_2+\bm{q}_3)-\phi (\bm{q}_1+\bm{q}_2+\bm{q}_3)]\rbrace\,,
\end{align}
\begin{align}
	&\textcolor{Silver Lake Blue}{p_5(\bm{q}_1,\bm{q}_2,\bm{q}_3,\bm{q}_5)+p_5(-\bm{q}_2,-\bm{q}_3,\bm{q}_5,\bm{q}_1)+p_5(\bm{q}_1,-\bm{q}_3,\bm{q}_6,\bm{q}_4)}\nonumber\\
	&\textcolor{Silver Lake Blue}{+p_5(-\bm{q}_6,\bm{q}_4,\bm{q}_3,\bm{q}_1)+p_5(-\bm{q}_2,\bm{q}_4,\bm{q}_6,\bm{q}_5)+p_5(-\bm{q}_6,\bm{q}_2,\bm{q}_5,\bm{q}_4)}=\nonumber\\
	&-36 N + 10[|\tilde{S}(\bm{q}_1+\bm{q}_2+\bm{q}_3)|^2+|\tilde{S}(\bm{q}_1+\bm{q}_2)|^2+|\tilde{S}(\bm{q}_1)|^2+ |\tilde{S}(\bm{q}_2+\bm{q}_3)|^2+ |\tilde{S}(\bm{q}_2)|^2+ |\tilde{S}(\bm{q}_3)|^2]\nonumber\\
	&+2[|\tilde{S}(\bm{q}_1+2
		\bm{q}_2+\bm{q}_3)|^2+ |\tilde{S}(\bm{q}_3-\bm{q}_1)|^2+
		|\tilde{S}(\bm{q}_1+\bm{q}_3)|^2]\nonumber\\
	&-6\lbrace |\tilde{S}(\bm{q}_2)| |\tilde{S}(\bm{q}_3)| |\tilde{S}(\bm{q}_2+\bm{q}_3)| \cos [\phi (\bm{q}_2)+\phi
		(\bm{q}_3)-\phi (\bm{q}_2+\bm{q}_3)]\nonumber\\
	&+ |\tilde{S}(\bm{q}_1)| |\tilde{S}(\bm{q}_2+\bm{q}_3)| |\tilde{S}(\bm{q}_1+\bm{q}_2+\bm{q}_3)| \cos [\phi (\bm{q}_1)+\phi
		(\bm{q}_2+\bm{q}_3)-\phi (\bm{q}_1+\bm{q}_2+\bm{q}_3)]\nonumber\\
	&+ |\tilde{S}(\bm{q}_3)| |\tilde{S}(\bm{q}_1+\bm{q}_2)| |\tilde{S}(\bm{q}_1+\bm{q}_2+\bm{q}_3)| \cos [\phi
		(\bm{q}_3)+\phi (\bm{q}_1+\bm{q}_2)-\phi (\bm{q}_1+\bm{q}_2+\bm{q}_3)]\nonumber\\
	&+ |\tilde{S}(\bm{q}_1)| |\tilde{S}(\bm{q}_2)| |\tilde{S}(\bm{q}_1+\bm{q}_2)| \cos [\phi (\bm{q}_1)+\phi (\bm{q}_2)-\phi (\bm{q}_1+\bm{q}_2)]\rbrace\nonumber\\
	&-2\lbrace |\tilde{S}(\bm{q}_1+\bm{q}_2)| |\tilde{S}(\bm{q}_3-\bm{q}_1)| |\tilde{S}(\bm{q}_2+\bm{q}_3)| \cos [\phi (\bm{q}_1+\bm{q}_2)+\phi (\bm{q}_3-\bm{q}_1)-\phi
		(\bm{q}_2+\bm{q}_3)]\nonumber\\
	&+ |\tilde{S}(\bm{q}_2)| |\tilde{S}(\bm{q}_1+\bm{q}_3)| |\tilde{S}(\bm{q}_1+\bm{q}_2+\bm{q}_3)| \cos [\phi
		(\bm{q}_2)+\phi (\bm{q}_1+\bm{q}_3)-\phi (\bm{q}_1+\bm{q}_2+\bm{q}_3)]\nonumber\\
	&+ |\tilde{S}(\bm{q}_1+\bm{q}_2)| |\tilde{S}(\bm{q}_2+\bm{q}_3)| |\tilde{S}(\bm{q}_1+2 \bm{q}_2+\bm{q}_3)| \cos [\phi (\bm{q}_1+\bm{q}_2)+\phi (\bm{q}_2+\bm{q}_3)-\phi
		(\bm{q}_1+2 \bm{q}_2+\bm{q}_3)]\nonumber\\
	&+ |\tilde{S}(\bm{q}_2)| |\tilde{S}(\bm{q}_1+\bm{q}_2+\bm{q}_3)| |\tilde{S}(\bm{q}_1+2
		\bm{q}_2+\bm{q}_3)| \cos [\phi (\bm{q}_2)+\phi (\bm{q}_1+\bm{q}_2+\bm{q}_3)-\phi (\bm{q}_1+2 \bm{q}_2+\bm{q}_3)]\nonumber\\
	&+ |\tilde{S}(\bm{q}_1)| |\tilde{S}(\bm{q}_3)| |\tilde{S}(\bm{q}_3-\bm{q}_1)| \cos [\phi (\bm{q}_1)-\phi (\bm{q}_3)+\phi (\bm{q}_3-\bm{q}_1)]\nonumber\\
	&+|\tilde{S}(\bm{q}_1)| |\tilde{S}(\bm{q}_3)| |\tilde{S}(\bm{q}_1+\bm{q}_3)| \cos [\phi (\bm{q}_1)+\phi (\bm{q}_3)-\phi (\bm{q}_1+\bm{q}_3)]\rbrace\nonumber\\
	&+2\lbrace |\tilde{S}(\bm{q}_1)| |\tilde{S}(\bm{q}_3)| |\tilde{S}(\bm{q}_1+\bm{q}_2)| |\tilde{S}(\bm{q}_2+\bm{q}_3)| \cos [\phi (\bm{q}_1)-\phi
		(\bm{q}_3)-\phi (\bm{q}_1+\bm{q}_2)+\phi (\bm{q}_2+\bm{q}_3)]\nonumber\\
	&+ |\tilde{S}(\bm{q}_1)| |\tilde{S}(\bm{q}_2)| |\tilde{S}(\bm{q}_3)| |\tilde{S}(\bm{q}_1+\bm{q}_2+\bm{q}_3)| \cos [\phi (\bm{q}_1)+\phi
		(\bm{q}_2)+\phi (\bm{q}_3)-\phi (\bm{q}_1+\bm{q}_2+\bm{q}_3)]\nonumber\\
	&+ |\tilde{S}(\bm{q}_2)| |\tilde{S}(\bm{q}_1+\bm{q}_2)| |\tilde{S}(\bm{q}_2+\bm{q}_3)| |\tilde{S}(\bm{q}_1+\bm{q}_2+\bm{q}_3)|\nonumber\\
	&\times \cos [\phi (\bm{q}_2)-\phi
		(\bm{q}_1+\bm{q}_2)-\phi (\bm{q}_2+\bm{q}_3)+\phi (\bm{q}_1+\bm{q}_2+\bm{q}_3)]\rbrace\,.
\end{align}
In total, we have
\begin{align}
	&N^4 g^{(4)}(\bm{q}_1,\bm{q}_2,\bm{q}_3)=N^4-12 N^3+60 N^2-144 N +(N^2-10N+32)\times\nonumber\\
	&\times[|\tilde{S}(\bm{q}_1)|^2+ |\tilde{S}(\bm{q}_2)|^2+|\tilde{S}(\bm{q}_3)|^2+|\tilde{S}(\bm{q}_1+\bm{q}_2)|^2+|\tilde{S}(\bm{q}_2+\bm{q}_3)|^2+|\tilde{S}(\bm{q}_1+\bm{q}_2+\bm{q}_3)|^2]\nonumber\\
	&+|\tilde{S}(\bm{q}_1)|^2 |\tilde{S}(\bm{q}_3)|^2+|\tilde{S}(\bm{q}_1+\bm{q}_2)|^2 |\tilde{S}(\bm{q}_2+\bm{q}_3)|^2+|\tilde{S}(\bm{q}_2)|^2 |\tilde{S}(\bm{q}_1+\bm{q}_2+\bm{q}_3)|^2\nonumber\\
	&+4[|\tilde{S}(\bm{q}_1+\bm{q}_3)|^2+|\tilde{S}(\bm{q}_1+2 \bm{q}_2+\bm{q}_3)|^2+|\tilde{S}(\bm{q}_3-\bm{q}_1)|^2]\nonumber\\
	&-4\lbrace|\tilde{S}(\bm{q}_1+\bm{q}_2)| |\tilde{S}(\bm{q}_3-\bm{q}_1)| |\tilde{S}(\bm{q}_2+\bm{q}_3)| \cos[\phi(\bm{q}_1+\bm{q}_2)+\phi(\bm{q}_3-\bm{q}_1)-\phi(\bm{q}_2+\bm{q}_3)]\nonumber\\
	&+|\tilde{S}(\bm{q}_2)| |\tilde{S}(\bm{q}_1+\bm{q}_3)| |\tilde{S}(\bm{q}_1+\bm{q}_2+\bm{q}_3)| \cos[\phi(\bm{q}_2)+\phi(\bm{q}_1+\bm{q}_3)-\phi(\bm{q}_1+\bm{q}_2+\bm{q}_3)]\nonumber\\
	&+|\tilde{S}(\bm{q}_1+\bm{q}_2)|
	|\tilde{S}(\bm{q}_2+\bm{q}_3)| |\tilde{S}(\bm{q}_1+2 \bm{q}_2+\bm{q}_3)| \cos[\phi(\bm{q}_1+\bm{q}_2)+\phi(\bm{q}_2+\bm{q}_3)-\phi(\bm{q}_1+2\bm{q}_2+\bm{q}_3)]\nonumber\\
	&+|\tilde{S}(\bm{q}_2)| |\tilde{S}(\bm{q}_1+\bm{q}_2+\bm{q}_3)| |\tilde{S}(\bm{q}_1+2 \bm{q}_2+\bm{q}_3)| \cos[\phi(\bm{q}_2)+\phi(\bm{q}_1+\bm{q}_2+\bm{q}_3)-\phi(\bm{q}_1+2\bm{q}_2+\bm{q}_3)]\nonumber\\
	&+|\tilde{S}(\bm{q}_1)| |\tilde{S}(\bm{q}_3)| |\tilde{S}(\bm{q}_3-\bm{q}_1)| \cos [\phi(\bm{q}_1)-\phi(\bm{q}_3)+\phi(\bm{q}_3-\bm{q}_1)]\nonumber\\
	&+|\tilde{S}(\bm{q}_1)| |\tilde{S}(\bm{q}_3)| |\tilde{S}(\bm{q}_1+\bm{q}_3)| \cos[\phi(\bm{q}_1)+\phi(\bm{q}_3)-\phi(\bm{q}_1+\bm{q}_3)]\rbrace\nonumber\\
	&+2(N-6)\lbrace|\tilde{S}(\bm{q}_3)||\tilde{S}(\bm{q}_1+\bm{q}_2)||\tilde{S}(\bm{q}_1+\bm{q}_2+\bm{q}_3)| \cos[\phi(\bm{q}_3)+\phi(\bm{q}_1+\bm{q}_2)-\phi(\bm{q}_1+\bm{q}_2+\bm{q}_3)]\nonumber\\
	&+|\tilde{S}(\bm{q}_1)| |\tilde{S}(\bm{q}_2+\bm{q}_3)| |\tilde{S}(\bm{q}_1+\bm{q}_2+\bm{q}_3)| \cos[\phi(\bm{q}_1)+\phi(\bm{q}_2+\bm{q}_3)-\phi(\bm{q}_1+\bm{q}_2+\bm{q}_3)]\nonumber\\
	&+|\tilde{S}(\bm{q}_1)| |\tilde{S}(\bm{q}_2)| |\tilde{S}(\bm{q}_1+\bm{q}_2)| \cos [\phi(\bm{q}_1)+\phi(\bm{q}_2)-\phi(\bm{q}_1+\bm{q}_2)]\nonumber\\
	&+|\tilde{S}(\bm{q}_2)| |\tilde{S}(\bm{q}_3)|
	|\tilde{S}(\bm{q}_2+\bm{q}_3)| \cos [\phi(\bm{q}_2)+\phi(\bm{q}_3)-\phi(\bm{q}_2+\bm{q}_3)]\rbrace\nonumber\\
	&+2\lbrace|\tilde{S}(\bm{q}_1)| |\tilde{S}(\bm{q}_3)| |\tilde{S}(\bm{q}_1+\bm{q}_2)| |\tilde{S}(\bm{q}_2+\bm{q}_3)| \cos[\phi(\bm{q}_1)-\phi(\bm{q}_3)+\phi(\bm{q}_2+\bm{q}_3)-\phi(\bm{q}_1+\bm{q}_2)]\nonumber\\
	&+|\tilde{S}(\bm{q}_1)| |\tilde{S}(\bm{q}_2)| |\tilde{S}(\bm{q}_3)| |\tilde{S}(\bm{q}_1+\bm{q}_2+\bm{q}_3)| \cos[\phi(\bm{q}_1)+\phi(\bm{q}_2)+\phi(\bm{q}_3)-\phi(\bm{q}_1+\bm{q}_2+\bm{q}_3)]\nonumber\\
	&+|\tilde{S}(\bm{q}_2)| |\tilde{S}(\bm{q}_1+\bm{q}_2)| |\tilde{S}(\bm{q}_2+\bm{q}_3)|
	|\tilde{S}(\bm{q}_1+\bm{q}_2+\bm{q}_3)|\times\nonumber\\ &\times\cos[\phi(\bm{q}_2)-\phi(\bm{q}_1+\bm{q}_2)-\phi(\bm{q}_2+\bm{q}_3)+\phi(\bm{q}_1+\bm{q}_2+\bm{q}_3)]\rbrace\,,
\end{align}
which can be used as an additional equation to check the validity of the phases of the structure factor reconstructed from the normalized third-order photon correlation function $g^{(3)}(\bm{q}_1,\bm{q}_2)$.
\end{widetext}


\bibliography{library}

\end{document}